\documentclass[twocolumn,epjc3]{svjour3}

\RequirePackage[T1]{fontenc}
\smartqed  

\RequirePackage{graphicx}
\RequirePackage{mathptmx}      
\RequirePackage{flushend}
\RequirePackage[numbers,sort&compress]{natbib}
\RequirePackage[colorlinks,citecolor=blue,urlcolor=blue,linkcolor=blue]{hyperref}

\journalname{Eur. Phys. J. C}

\usepackage{graphics}
\usepackage{xspace}
\usepackage{tabularx}
\usepackage{booktabs}
\usepackage{multirow}
\usepackage{subfig}
\usepackage{ifthen}
\usepackage{marginnote}
\usepackage{color}
\usepackage{soul}

\newcommand{\Gerda}{{\sc Gerda}\xspace}
\newcommand{\LArGe}{LArGe\xspace}
\newcommand{\Qbb}{$Q_{\beta\beta}$\xspace}
\newcommand{\znbb}{$0\nu\beta\beta$\xspace}
\newcommand{\tnbb}{$2\nu\beta\beta$\xspace}
\newcommand{\Roi}{{\sc Roi}\xspace}
\newcommand{\LNGS}{{\sc Lngs}\xspace}
\newcommand{\MPIK}{{\sc Mpik}\xspace}

\newcolumntype{C}[1]{>{\centering\arraybackslash}p{#1}} 
\newcommand{\forloop}[5][1]{%
\setcounter{#2}{#3}%
\ifthenelse{#4}{#5\addtocounter{#2}{#1}%
\forloop[#1]{#2}{\value{#2}}{#4}{#5}}{}}
\newcounter{crcounter}
\newcommand{\compensaterule}[1]{%
\forloop{crcounter}{1}{\value{crcounter} < #1}%
{\vspace*{-\aboverulesep}\vspace*{-\belowrulesep}}}
\newcommand{\multirowbt}[3]{\multirow{#1}{#2}%
{\compensaterule{#1}#3}}

\begin{document}

\title{\LArGe $ $-- Active background suppression using argon scintillation for the \Gerda \znbb-experiment}

\author{M Agostini\thanksref{addr1} \and M Barnab\'{e}-Heider\thanksref{addr1,addr2} \and D Budj\'{a}\v{s}\thanksref{addr1} \and C Cattadori\thanksref{addr3} \and A Gangapshev\thanksref{addr2,addr4} \and K Gusev\thanksref{addr1,addr5,addr6} \and M Heisel\thanksref{addr2,e1} \and M Junker\thanksref{addr7} \and A Klimenko\thanksref{addr2,addr5} \and A Lubashevskiy\thanksref{addr2,addr5} \and K Pelczar\thanksref{addr8} \and S Sch\"onert\thanksref{addr1} \and A Smolnikov\thanksref{addr2} \and G Zuzel\thanksref{addr2,addr8}
}

\thankstext{e1}{e-mail: mark.heisel@mpi-hd.mpg.de}

\institute{Technische Universit\"at M\"unchen, Germany\label{addr1} \and Max-Planck-Institut f\"ur Kernphysik, Heidelberg, Germany\label{addr2} \and Universit\`{a} degli Studi di Milano e INFN, Milano, Italy\label{addr3} \and Institut for Nuclear Research, Moscow, Russia\label{addr4} \and Joint Institut for Nuclear Research, Dubna, Russia\label{addr5} \and National Research Center Kurchatov Institut, Moscow, Russia\label{addr6} \and Laboratori Nazionali del Gran Sasso, Assergi, Italy\label{addr7} \and Jagellonian University, Cracow, Poland\label{addr8}}

\date{Received: date / Revised version: 27/05/15}

\maketitle

\abstract{
\LArGe is a \Gerda low-background test facility to study novel background suppression methods in a low-background environment, for future application in the \Gerda experiment. Similar to \Gerda, \LArGe operates bare germanium detectors submersed into liquid argon (1\,m$^3$, 1.4\,tons), which in addition is instrumented with photomultipliers to detect argon scintillation light. The scintillation signals are used in anti-coincidence with the germanium detectors to effectively suppress background events that deposit energy in the liquid argon. The background suppression efficiency was studied in combination with a pulse shape discrimination (PSD) technique using a BEGe detector for various sources, which represent characteristic backgrounds to \Gerda. Suppression factors of a few times $10^3$ have been achieved. First background data of \LArGe with a coaxial HPGe detector (without PSD) yield a background index of \mbox{(0.12$-$4.6)$\cdot 10^{-2}$} \mbox{cts/(keV$\cdot$kg$\cdot$y)} (90\% C.L.), which is at the level of \Gerda \mbox{Phase I}. Furthermore, for the first time we monitor the natural $^{42}$Ar abundance (parallel to \Gerda), and have indication for the \tnbb-decay in natural germanium. These results show the effectivity of an active liquid argon veto in an ultra-low background environment. As a consequence, the implementation of a liquid argon veto in \Gerda \mbox{Phase II} is pursued.
\PACS{
      {23.40.-s} {$\beta$ decay; double $\beta$ decay; electron and muon capture} \and
      {27.50.+e} {mass $59\leq A\leq89$} \and
      {29.30.Kv} {X- and $\gamma$-ray spectroscopy} \and
      {29.40.Mc} {Scintillation detectors}
     } 
} 


\section{Introduction}

\Gerda is an experiment to search for the neutrinoless double beta (\znbb ) decay of $^{76}$Ge. Bare high-purity germanium (HPGe) detectors enriched in $^{76}$Ge, which serve both as source and detector for the \znbb-decays, are submersed in liquid argon (LAr).  The LAr serves as a high purity shield against external radiation and as a coolant for the HPGe detectors. The \znbb-signal is a sharp peak in the energy spectrum at \Qbb$=$ 2039\,keV from the sum energy of the two beta particles in a single HPGe detector. Details of the experimental setup and performance are summarized in \cite{Gerda13_instrumentation}. The \Gerda experiment follows a staged approach: \mbox{Phase I} has been recently completed after acquiring an exposure of 21.6\,kg$\cdot$yr and a background count rate at \Qbb of \mbox{1$\cdot$10$^{-2}$} \mbox{cts/(keV$\cdot$kg$\cdot$yr)} after pulse shape analysis \cite{Gerda13_psd,Gerda_background}. No signal was observed and a limit for the half-life of ${T^{0\nu}_{1/2}} > 2.1 \cdot 10^{25}$\,yr (90\% C.L.) was achieved \cite{Gerda13_PRL}. \mbox{Phase II} is currently under preparation: the goal is to explore half-live values in the range of $10^{26}$\,yr by further reducing the background by one order of magnitude to \mbox{$\le$10$^{-3}$} \mbox{cts/(keV$\cdot$kg$\cdot$yr)}, and by collecting an exposure of up to 100\,kg$\cdot$yr quasi background free.

\subsection{Background suppression in \Gerda Phase II}

To reach this demanding background count rate, several experimental modifications with respect to \mbox{Phase I} are being implemented: the most important are (1) the additional deployment of approximately 20\,kg novel thick-window Broad-Energy Germanium (BEGe) detectors with highly efficient pulse shape discrimination (PSD) performance \cite{Budjas09b,Agostini11}, and (2) the implementation of a sensor system to detect the liquid argon scintillation light in anti-coincidence with the germanium detectors for background suppression, first demonstrated in references \cite{Peiffer05,DiMarco07}.

\LArGe, the {\underline L}iquid {\underline {Ar}}gon {\underline {Ge}}rmanium test facility of \Gerda , was constructed to study these novel active background suppression methods in a low-background environment \cite{MiniLargePaper,Heisel11}. Similar to \Gerda , bare Ge-detectors are operated in \LArGe in 1\,m$^3$ (1.4\,tons) of liquid argon, which in addition is instrumented with photomultiplier tubes (PMT). The setup is located underground at 3800\,m w.e. at the Laboratori Nazionali del Gran Sasso (\LNGS), and has been taking data since May 2010. The data presented in this paper demonstrates that the argon scintillation veto technique works very efficiently both alone and in combination with PSD applied to BEGe detector signals.

\subsection{Concept of liquid argon scintillation veto}

It is well known that liquid argon emits scintillation light in response to ionizing radiation. Up to approximately 40000 photons, peaked at a wavelength of 128\,nm (XUV photons), are emitted per 1\,MeV beta/gamma energy deposition \cite{Miyajima74}. Background events often have energy deposition outside the Ge-detector in the surrounding medium, in our case the scintillating LAr. Conversely, \znbb-events are confined to the Ge-detector, so that no scintillation light is triggered. An observation of the light is therefore a good indicator for a background event, and can be used to veto the coincident Ge-signal. For that purpose, the LAr must be instrumented to detect the light. In case of \LArGe the light is shifted in its wavelength to match the sensitive range of the PMTs and is guided to the PMTs with mirror foil on the inner cryostat wall.

\section{Experimental}

\subsection{Detector design}

\begin{figure}
\begin{minipage}{0.23\textwidth}
\vspace{5ex}
\textbf{lock}\\
for Ge-detector deployment\\
\vspace{-0.8ex}\\
\textbf{copper cryostat}\\
with WLS mirror foil\\
LAr volume 1\,m$^3$ (1.4\,t)\\
\vspace{-0.8ex}\\
\textbf{PMTs}\\
9$\times$ 8" ETL 9357\\
coated with WLS\\
\vspace{-0.8ex}\\
\textbf{detector strings}\\
up to 9 Ge-detectors\\
\vspace{-0.8ex}\\
\textbf{graded shield}\\
15\,cm copper, 10\,cm lead,\\
23\,cm steel, 20\,cm polyethylene\\
\end{minipage}
\begin{minipage}{0.24\textwidth}
\resizebox{1\textwidth}{!}{
	\includegraphics{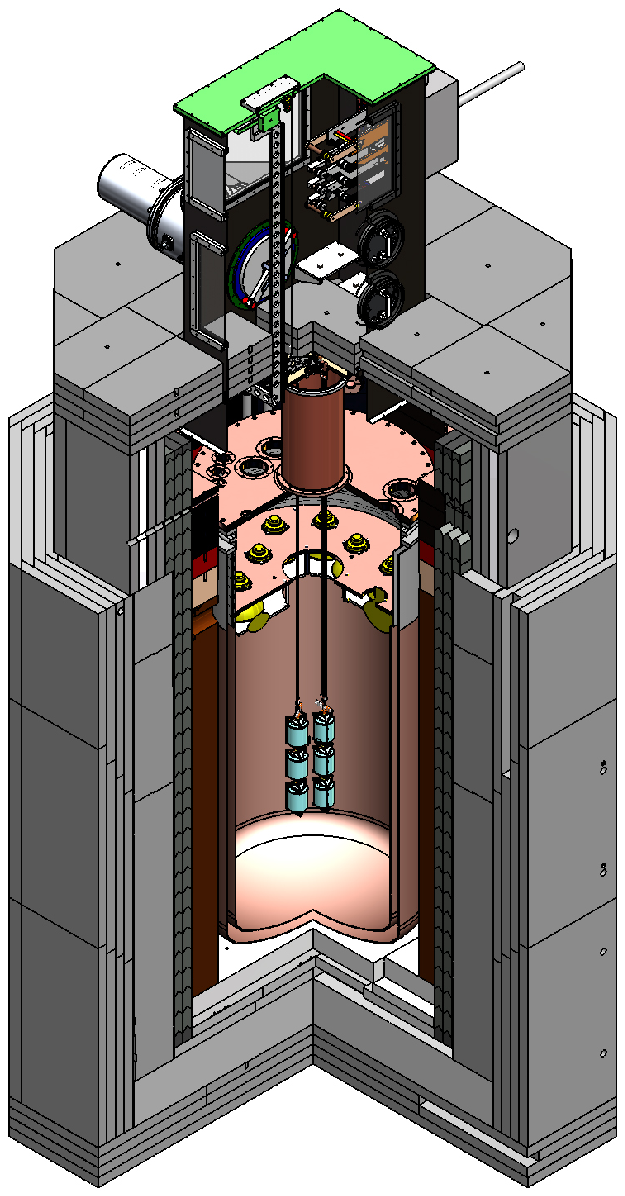}}
\end{minipage}
\begin{picture}(0,0)(241,95)
   \put(5,151){\line(1,0){128}}
   \put(5,114){\line(1,0){119}}
   \put(5,77){\line(1,0){119}}
   \put(5,50){\line(1,0){119}}
   \put(5,13){\line(1,0){119}}  
   \thicklines
   \color{white}
   \put(133,151){\vector(4,0){30}}
   \put(124,114){\vector(4,0){40}}
   \put(124,77){\vector(3,1){45}}
   \put(124,50){\vector(3,1){55}}
   \put(124,13){\vector(3,1){25}}
\end{picture}
\caption{Cutaway view of the \LArGe setup.}
\label{fig:LArGe3D}
\end{figure}

An illustration of the \LArGe setup is shown in Fig. \ref{fig:LArGe3D}. A vacuum insulated copper cryostat at the center can hold 1000\,l (1.4\,t) of LAr. Nine 8" ETL 9357 photomultiplier tubes are immersed into the LAr from the top. The inner cryostat walls are lined with VM2000 radiant mirror foil\footnote{product of the company 3M}. PMTs and mirror foil are coated with TPB/polysterene wavelength shifter. The cryostat is surrounded by a passive shield against external background. A double chamber lock system on the top serves as an access port for the deployment of Ge-detectors and internal radioactive sources for calibration. The cryogenic infrastructure, a slow control system, and the DAQ are located adjacent to this setup.

\subsubsection{Shielding}

The graded shield of increasing radiopurity is designed to have the gamma background dominated by the innermost layer of 15\,cm electrolytic copper. This is followed by 10\,cm of low-activity lead and 23\,cm of steel. The outermost layer consists of 20\,cm polyethylene to attenuate neutrons. Together, all three layers attenuate an external gamma-ray from the 2615\,keV line of $^{208}$Tl to 5$\cdot 10^{-8}$ of the initial flux. The purest shielding is provided by $>$40\,cm LAr inside the cryostat. At the inside of the cryostat, the PMT glass and the VM2000 mirror foil yield the highest radioimpurities. To maintain sufficient shielding, the distance of the PMTs to the Ge-detectors is chosen as large as 90\,cm. For the mirror foil, the low area density leads to a surface activity below that of the copper. The radiopurity screening results of these \LArGe components are given in Table \ref{tab:exp_shield}. Throughout the measurements presented here, a part of the steel and polyethylene shield was not completed yet. Only a small impact on the background run is expected thereof.

\begin{table*}
\begin{center}
\caption{Gamma-spectrometry measurements of radioimpurities in \LArGe materials. Given is the specific activity in units [mBq/kg]. The origin of the materials is; Copper {\sc Lens}: NOSV copper from NAA Hamburg. Lead {\sc Lens}: `Doe Run'-quality from JL Goslar. Steel (shield): carbon steel from Dillinger H\"utte GTS. PE (shield): polyethylene from Simona AG. VM2000: radiant mirror foil from 3M (measured with glue, mounted with glue removed). PMT glass (bulb): ultra-low background glass from ETL (now ET Enterprise Ltd.).}
	\label{tab:exp_shield}
    \begin{tabular}{lcccccc}
    \toprule material & \multicolumn{6}{c}{specific activity} \\
    & $^{226}$Ra & $^{228}$Ra & $^{228}$Th & $^{40}$K & $^{60}$Co & others \\
	\midrule copper {\sc Lens}$^{a,1}$ & $<$0.016   &          & $<$0.019 & $<$0.088 & $<$0.01 	& \\
	\midrule lead {\sc Lens}$^{a,1}$   & $<$0.029   &          & $<$0.022 & 0.44(14) & 0.18(2) 	& 27(4)$\cdot10^3$ $^{210}$Pb \\ 
	\midrule steel (shield)$^{b,2}$    & 2.04(33)   & 1.63(41) & 5.34(69) & $<$4.2   & $<$0.3 	& \\
	\midrule PE (shield)$^{b,2}$       & $<$2.5     & 11.2(32) & $<$3.4   & 10.8(67) & 			& \\
	\midrule VM2000$^{a,2}$	           	& $<$1.6     & $<$2.2   & $<$1.2   & 140(10)  & $<$0.44 	& $<$0.45 $^{137}$Cs \\
	\midrule PMT glass (bulb)$^{b,2}$  & 2010(190)  & 2010(480)& 210(60)  & 1750(430)& 			& $<$370 $^{137}$Cs \\
	\bottomrule 
	\multicolumn{3}{l}{{\footnotesize $^a$measured with the {\sc Gempi} spectrometer at \LNGS. \cite{Heusser04}}} &
	\multicolumn{3}{l}{{\footnotesize $^1$upper limits given for k=2, 97.7\% C.L.}} \\
	\multicolumn{3}{l}{{\footnotesize $^b$measured in the Low-Level Lab at \MPIK Heidelberg. \cite{Heusser90}}} &
	\multicolumn{3}{l}{{\footnotesize $^2$upper limits given for k=1.645, 95\% C.L.}} \\
  \end{tabular}
\end{center}
\end{table*}

\subsubsection{Cryostat \& cryogenics}

The main body of the double-wall cryostat is made from electrolytic copper. It has an inner diameter of 90\,cm and a height of 210\,cm. Heat loss is primarily prevented by the insulation vacuum of 10$^{-5}$ to 10$^{-6}$\,mbar. The uppermost 40\,cm collar and bellow are made from stainless steel to minimize the thermal conductivity to the top. The cryostat is closed by a 38\,mm thick flange from electrolytic copper. Infrared shields from thin copper foil are mounted both inside the insulation vacuum, and below the top flange. A PMT support structure of copper and PTFE rests on a rim at the copper-steel transition. The filling level of the LAr is adjusted slightly above the PMTs' equator, leaving the upper part of the cryostat in the gas phase of argon. A strong temperature gradient builds up from the copper at LAr temperature, across the stainless steel to the top flange, which itself stays above the freezing point. The total heat load is $\sim$90\,W.

An active cooling system cools the cryostat by evaporating liquid nitrogen (LN$_2$) in an integrated cooling spiral in the steel collar. The innermost infrared shield (3\,mm thick) is thermally coupled to this heat exchanger to prevent heat lost through radiation. All relevant cryogenic parameters are compiled in a slow control system, which regulates the cooling power via the flow of LN$_2$. During normal operation a LN$_2$ flow of 2.5\,kg/h is sufficient to reduce the LAr loss to zero, which allows a continuous operation of \LArGe . The working pressure of the cryostat is kept at 30$-$70\,mbar overpressure, to prevent gaseous impurities from the outside to enter the LAr.

The filling procedure for the cryostat requires special precaution to prevent contamination of the LAr. The measurements presented in this article were performed in LAr 5.5 (purity 99.9995\%). Against traces of humidity several pumping-flushing cycles were performed with gaseous argon, while the cryostat was heated to $>$40$^\circ$C. To prevent radioactive background from radon, the argon was filled through an active-charcoal trap (602\,g of CarboAct), followed by a PTFE particle filter.

\subsubsection{Lock \& source insertion system}
\label{sec:lock}

The lock on top of the \LArGe setup serves for the deployment of Ge-detector strings and internal radioactive wire sources into the cryostat. So far, only one detector has been inserted at a time. The Ge-detectors are mounted to low-mass copper holders in a separate cleanbench. They are transferred to the main lock using a transportation container, keeping them in gaseous nitrogen atmosphere at all times. The main lock and the open volume between cryostat and shield are permanently flushed with gaseous argon at slight overpressure. The pressure gradient from the cryostat, via lock and shield, to the outside acts as a safeguard against air-trace contamination of the LAr and avoids background from airborne $^{222}$Rn.

The Ge-detectors are deployed into the cryostat approximately equally spaced between the bottom and side walls (distance d$\,\approx\,$45\,cm), leaving 90\,cm of LAr on top towards the PMTs. Internal `close-by' wire sources are inserted directly adjacent to the Ge-diodes in the argon (d$\,\approx\,$7\,cm). External sources are deployed through access ports in the shield directly adjacent to the outer wall of the cryostat (d$\,\approx\,$50\,cm). The vertical position of all sources matches the center of the Ge-detector.

\begin{figure}
\begin{center}
\resizebox{0.485\textwidth}{!}{%
  \includegraphics{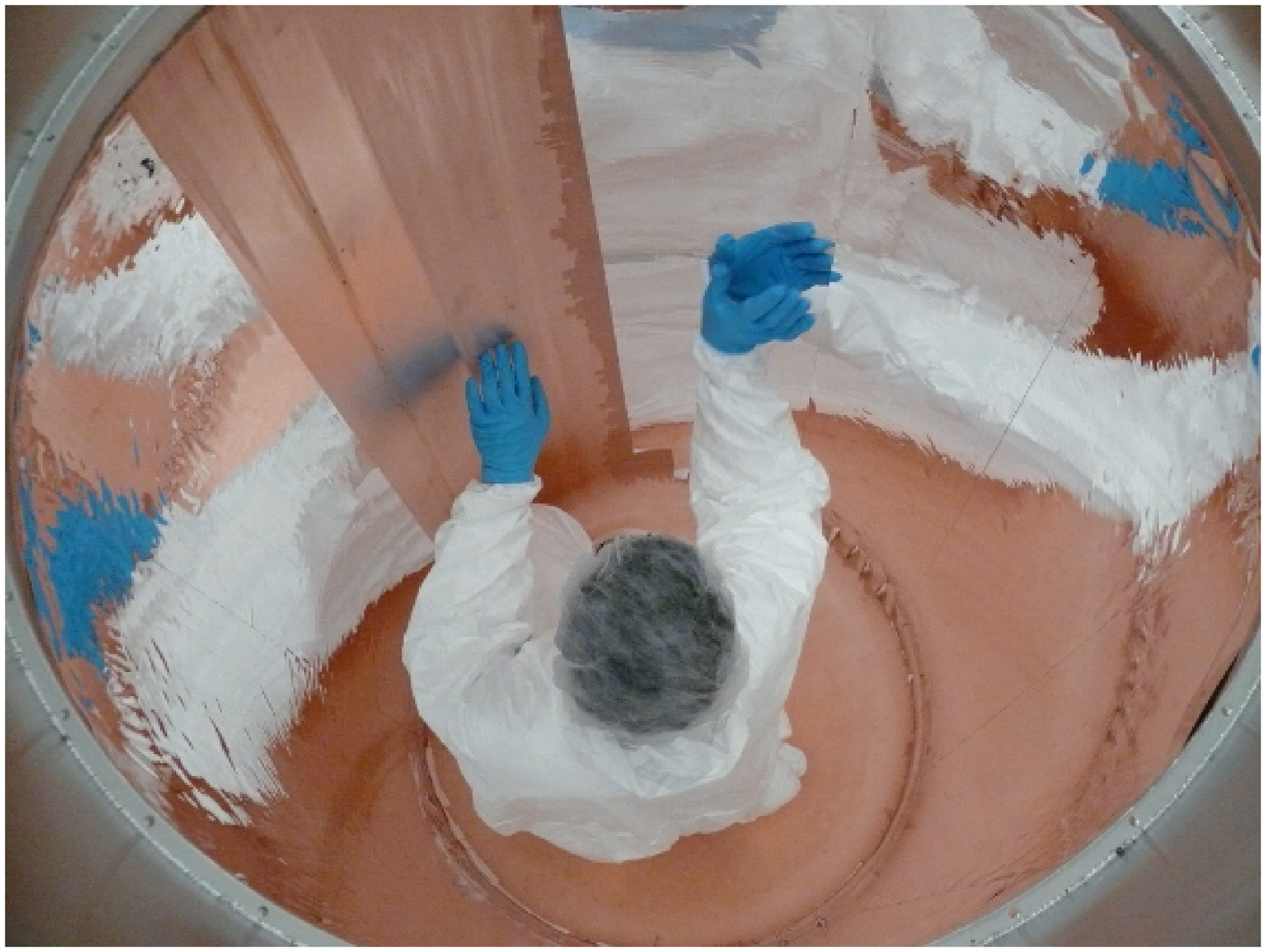}
  \hspace{1ex}
  \includegraphics{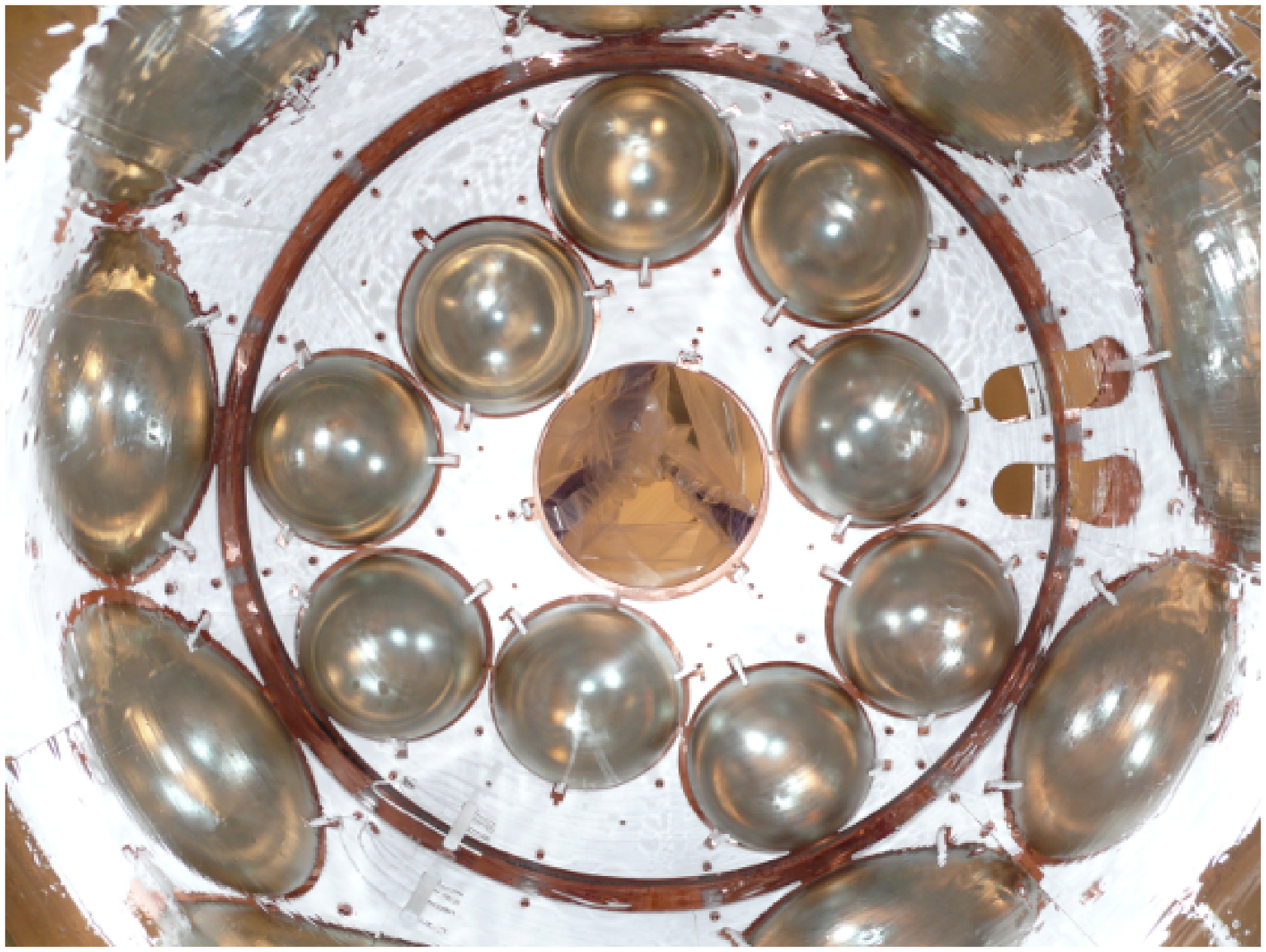}}
\end{center}
\caption{\textbf{Left:} Cryostat beeing lined with VM2000 mirror foil. \textbf{Right:} Bottom view of the PMTs inside the cryostat.}
\label{fig:inside_cryostat}
\end{figure}

\subsection{Light instrumentation}
\label{sec:instrumentation}

The light read-out is done using nine ETL\footnote{now: ET Enterprise Ltd.} 9357 photomultiplier tubes. The 8" (200\,mm) diameter end window with a low resistance bialkali photocathode is sensitive to wavelengths from 275$-$630\,nm\footnote{wavelength range over which quantum efficiency exceeds 1\%, according to the data sheet.}. The peak quantum efficiency of the PMTs is 18\% at 370\,nm. Since the glass of the end window is not transparent for the 128\,nm scintillation light, it must be covered with wavelength shifter (WLS). A picture of the PMTs and mirror foil in the setup is shown in Fig. \ref{fig:inside_cryostat}.

The PMTs are equipped with a custom made voltage divider with a wide linear dynamic range from 2\,mV to 4\,V. Clean pulse shapes are obtained by operating with negative HV on the photocathode. The voltage divider is based on a 0.5\,mm thin CuFlon{\footnotesize \textregistered} printed circuit board with components selected for low mass and radiopurity.

The wavelength shifter used to coat Mirror foil and PMTs consists of 10\% fluorescent dye (tetraphenyl butadiene) embedded into a polymer matrix of purified poly\-sty\-rene. Both substances are dissolved in toluene and dehumidified using gaseous nitrogen. A homogeneous coating is achieved by pulling the Mirror foil through a bath in 45$^\circ$ angle. In case of PMTs, the photocathodes are simply brushed with the solution in several thin layers. The coating thickness is estimated from the consumption of the solution. A thickness of 1$-$4\,$\mu$m is choosen as a compromise between shifting efficiency and mechanical stability in cryogenic liquid. The coated foil has a specular reflectivity of $\sim$95\% at the peak fluorescence wavelength around 420\,nm.

\begin{figure}
	\vspace{-7mm}
	\begin{minipage}[h!]{0.272\textwidth}	
		\resizebox{1\textwidth}{!}{
		\includegraphics{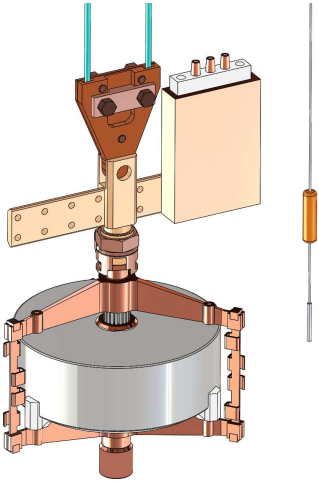}}
		\vspace{-1cm}
	\end{minipage}
	\hfill
	\begin{minipage}[h!]{0.205\textwidth}	
		\vspace{21mm}
		\caption{Schematic view of the BEGe detector (grey colour) in a low-mass copper-holder. The CC2 preamplifier is placed in a grounded copper box mounted above the detector (top right). No electronic cabling is shown. A close-by wire source with copper weight is hanging on the right side.}
		\label{fig:BEGe}
	\end{minipage}
	\vspace{-4mm}
\end{figure}

A characteristic quantity of the setup is the photoelectron (pe) yield $Y$. For the measurements discussed here we had $Y\approx0.05$\,pe/keV, which corresponds to an energy threshold of $1/Y\approx20$\,keV for a single photoelectron. Two reasons for this low $Y$ are: (1) only 5 out of 9 PMTs were operational at low temperature\footnote{Some PMTs exhibit light production due to discharges within the glass body at cryogenic temperatures and were either operated at reduced gain or not at all. These findings led to an R\&D program to solve this issue for next generation PMTs being implemented in \Gerda Phase II.}, (2) trace contaminations of the argon can strongly quench the scintillation and significantly shorten the attenuation length of the 128\,nm photons. An indicator for the light quenching is the lifetime $\tau$ of the triplet states of the argon excimers, which can be measured from recorded scintillation waveforms. During data taking $\tau$ was monitored in the range between 450 to 550\,ns, compared to $\tau=1590$\,ns \cite{HTF83} in clean argon.

\subsection{Germanium detectors}
\label{sec:Germanium detectors}

A modified thick-window broad-energy germanium (BEGe) detector has been used to carry out the suppression-ef\-fi\-cien\-cy measurements described in section \ref{sec:suppression}. It is a p-type diode of 878\,g by Canberra Semiconductors, N.V. Olen/Belgium. The depletion voltage is +4\,kV \cite{Budjas09a}. A small $p+$ contact leads to a strong weighting field close to the read-out electrode, which allows good pulse shape discrimination. A prototype of the \Gerda \mbox{Phase I} multi-channel charge-sensitive preamplifier (CC2) with integrated FET and RC feedback component \cite{Riboldi10} is used and mounted close-by the detector -- see Fig. \ref{fig:BEGe}. A low-mass copper-holder is used to support the diode and submerse it `naked' into LAr. Signal and HV are connected via pressure contacts to the detector surface. A detector re\-so\-lu\-tion of 1.99\,keV FWHM at 1332\,keV is achieved in this setup, compared to 1.63\,keV in a vacuum cryostat with the same detector \cite{Budjas09b}.

For the background measurements described in section \ref{sec:background} the BEGe detector is replaced by the coaxial p-type HPGe-diode GTF44. In contrast to BEGe it has a low intrinsic background ($^{60}$Co, $^{68}$Ge) and a high mass (2465\,g), whereas pulse shape discrimination is inferior and not applied here. The detector has been modified by the manufacturer Canberra for the bare operation in LAr \cite{heider09}. It is equipped with a low-background version of the CC2 charge sensitive preamplifier. Within an investigation program of $^{42}$Ar background the diode has been encapsuled in a grounded Faraday cage made from thin layers of PTFE and copper.

\subsection{Data aquisition and analysis methods}

\subsubsection{Combined Ge-detector and PMT readout \& waveform processing}

\begin{figure}
	\resizebox{0.48\textwidth}{!}{
	\includegraphics{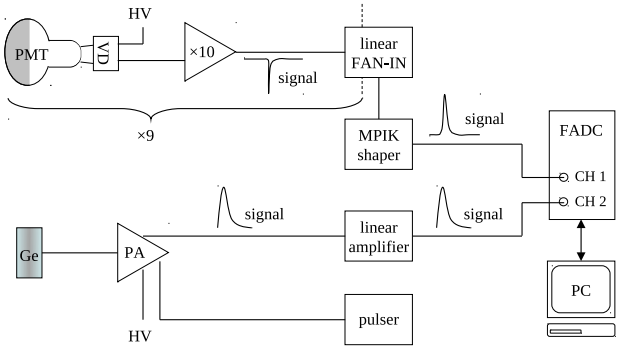}}
	\vspace{-1.8cm}
	\caption{Block diagram of the DAQ with Ge-detector and PMT readout. VD = voltage divider; PA = preamplifier.}
	\label{fig:electronic}
\end{figure}

A block diagram of DAQ and front-end electronics is shown in Fig. \ref{fig:electronic}. The Ge-detector is supplied with bias HV (Iseg NHQ 225M NIM). A pulser signal (Ortec Mod. 448 NIM) can be fed into a test input of the preamplifier. The output signal is amplified without shaping by a custom-made linear amplifier, and fed into a FADC (Struck SIS3301 VME; 8 channels, 14-bit, 105\,MS/s). The PMT HV is supplied by Iseg NHQ 204M/225M. The signals are amplified by a factor ten (Phillips Scientific Mod. 776 NIM) and merged in a linear fan-in (LeCroy Mod. 428F NIM). The resulting analog-sum is amplified in another custom-made analog shaper (NIM) with a shaping constant of a few 10\,ns, to match the dynamic range and sampling rate of the subsequent FADC. The FADC is internally triggered on the Ge-signal, and simultaneously records Ge- and PMT waveforms of 40$\,\mu$s trace-length with 100\,MHz sampling rate. The FADC aquisition is controlled by a custom-made software by MIZZI Computer Software GmbH \cite{Mizzi}.

The offline analysis of the digitized Ge-waveforms is performed with the software framework {\sc Gelatio} \cite{Agostini11_2} and following the procedure described in reference \cite{Agostini12}. The energy deposited in the Ge-detectors is reconstructed by applying an approximated Gaussian digital filter. Non-physical signals due to instabilities of the read-out electronics or electromagnetic pick-up noise are rejected during the data processing by applying a sequence of quality cuts; including the flatness and noise-level of the baseline, the position and rise time of the leading edge, and the fall time of the exponential decay tail of the charge signal. The PMT waveforms are analysed without filtering or quality cuts. Merely the baseline and veto condition are determined.

\subsubsection{LAr scintillation veto cut}
\label{sec:larvetocut}

The veto condition is fulfilled when one or more photoelectrons are detected in a 5$\,\mu$s window around the Ge-trigger. Threshold and window size are optimized to maximize the product of suppression factor $SF$ at \Qbb and the veto acceptance $\epsilon_{acc}$. The threshold is slightly above noise at $5\sigma$ of the baseline spread, corresponding to $\sim$20\% of the average single photoelectron amplitude.

The veto acceptance is the complementary probability for an event being vetoed by random coincidences ($p_{rc}$), \mbox{$\epsilon_{acc}=1-p_{rc}$}. It is measured by applying the veto cut on pulser signals or single full energy peaks. The random coincidence probability can also be estimated via \mbox{$p_{rc}\approx\nu^{PMT}_{trig}\cdot\Delta t$}, using the PMT trigger rate and the veto window size. It turns out that all methods yield consistent results. Veto acceptance values in measurements with different sources are listed in Table \ref{tab:sources}.

\begin{table}
\begin{center}
	\caption{\label{tab:sources} Summary table of source activities and measurement time with corresponding signal rates of Ge-detector, PMTs, and the pulser acceptance.}	
  	\begin{tabular}{ccccccc}
    \toprule source & pos. & activity & meas. & $\nu^{BEGe}_{trig}$ & $\nu^{PMT}_{trig}$ & $\epsilon_{acc}$ \\
											&     & [kBq] 	& time [d] 	& [Hz]	& [kHz]	& [\%] \\
    \midrule $^{60}$Co  				  	& int & 0.314 	& 13.06 	& 26.9 	& 5.8 	& 96.6 \\
    \midrule \multirowbt{2}{*}{$^{226}$Ra}	& ext & 94.9 	& 2.49 		& 6.9 	& 39.2 	& 78.7 \\
    \cmidrule{2-7}                        	& int & 0.934 	& 1.54 		& 65.4 	& 8.1 	& 94.2 \\
    \midrule \multirowbt{2}{*}{$^{228}$Th}	& ext & 38.9 	& 8.30 		& 4.5 	& 43.3 	& 78.3 \\
    \cmidrule{2-7} 						  	& int & 0.63 	& 1.96 		& 47.5 	& 8.6 	& 95.7 \\
    \bottomrule 
  \end{tabular}
  \vspace{-1.8em}
\end{center}
\end{table}

The suppression factor is the ratio of events in the unsuppressed ($N_0$) versus the suppressed ($N_S$) spectrum. To make $SF$ independent of the source strength in the measurement it is weighted by $\epsilon_{acc}$, hence \mbox{$SF=\epsilon_{acc}\cdot N_0/N_S$}. Suppression factors for the \Roi are determined in a 70\,keV window around \Qbb. Uncertainties are calculated according to Poisson counting statistics.

\subsubsection{Pulse shape discrimination}

The objective of PSD is to distinguish the \emph{single site events} (SSE) of the $\beta\beta$-decay from \emph{multi site events} (MSE) of common gamma-background with multiple interaction vertices within the Ge-diode. SSE and MSE can be distinguished by the maximum amplitude (A) of their current pulse, normalized by the energy (E). The so-called cut parameter 'A/E' has been established for this purpose \cite{Budjas09b}. The PSD cut is calibrated to 90\% acceptance on the double escape peak (DEP) of the 2615\,keV $^{208}$Tl line, which by its nature is dominantly of SSE character. As discussed in \cite{Gerda13_psd}, in a BEGe detector the DEP is a good proxy for \znbb and their acceptances agree within about 1\%. Uncertainties of PSD suppression factors include both statistical and systematic uncertainties.

In addition, the combined suppression of LAr veto and PSD is determined from the spectra. Due to the strong suppression of close-by $^{60}$Co and $^{228}$Th the analysis window for these sources is extended from 70\,keV width to 200\,keV, excluding the single escape peak of $^{208}$Tl at 2104\,keV.

\begin{figure*}
	\hspace{1cm}\textbf{\subfloat[\textnormal{$^{228}$Th source close-by (top) \& external (bottom)}]{
	\hspace{-1cm}\footnotesize{\textnormal{\label{fig:Th228_full}\input{}}}}}
	\vspace{1.6mm}
	
	\hspace{1cm}\textbf{\subfloat[\textnormal{$^{226}$Ra source close-by (top) \& external (bottom)}]{
	\hspace{-1cm}\footnotesize{\textnormal{\label{fig:Ra226_full}\input{}}}}}
	\vspace{1.6mm}
	
	\hspace{1cm}\textbf{\subfloat[\textnormal{$^{60}$Co source close-by}]{
	\hspace{-1cm}\footnotesize{\textnormal{\label{fig:Co60_full}\input{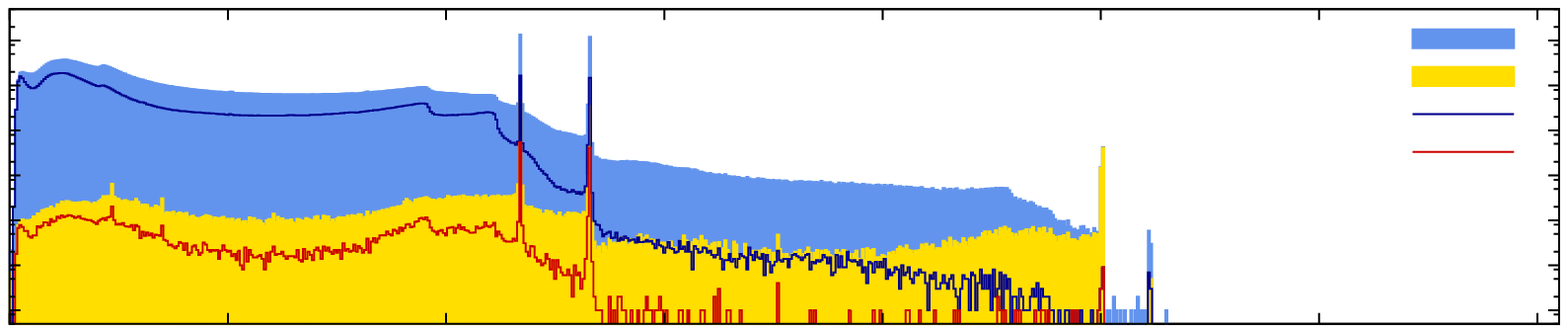}}}}}
	\vspace{1.6mm}
	
\caption{Full energy spectra of the source measurements in 5\,keV binning. The $^{228}$Th and $^{226}$Ra spectra feature a pulser signal at 3\,MeV. In case of $^{60}$Co, a small background peak of $^{208}$Tl is visible at 2615\,keV. The colour code is shown in canvas (c).}
\label{fig:full}
\end{figure*}

\begin{figure}
	\hspace{1cm}\textbf{\subfloat[\textnormal{$^{228}$Th source close-by (top) \& external (bottom)}]{
	\hspace{-1cm}\footnotesize{\textnormal{\label{fig:Th228_ROI}\input{}}}}}
	\vspace{1.6mm}
	
	\hspace{1cm}\textbf{\subfloat[\textnormal{$^{226}$Ra source close-by (top) \& external (bottom)}]{
	\hspace{-1cm}\footnotesize{\textnormal{\label{fig:Ra226_ROI}\input{}}}}}
	\vspace{1.6mm}

	\hspace{1cm}\textbf{\subfloat[\textnormal{$^{60}$Co source close-by}]{
	\hspace{-1cm}\footnotesize{\textnormal{\label{fig:Co60_ROI}\input{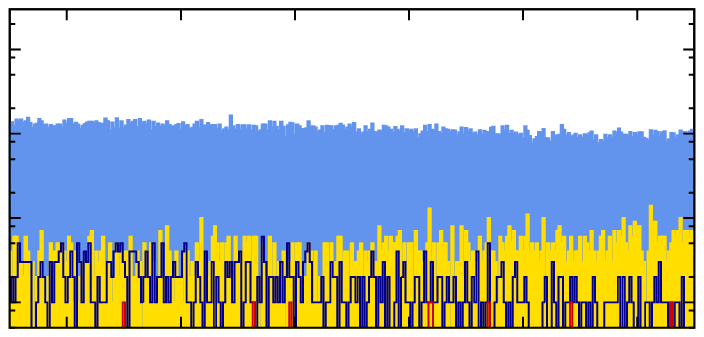}}}}}
	\vspace{1.6mm}
	
\caption{Close-up view of the {\Roi} at $Q_{\beta\beta}$ (2039\,keV) of the source spectra in 1\,keV binning. For colour code see Fig. \ref{fig:full}.}
\label{fig:ROI}
\end{figure}

\section{Measurement of suppression factors}
\label{sec:suppression}

Energy spectra of various sources in \emph{close-by} and \emph{external} position (see \ref{sec:lock}) were recorded with the BEGe detector. The sources represent characteristic background contributions in \Gerda . While from external sources only gammas can enter the cryostat, close-by sources are encapsuled in ceramics ($D=1\,$mm) and thin steel ($R=0.25-0.5\,$mm), thus allow some high-energy beta particles to enter the LAr as well. The source activities are chosen to balance high signal rates in the Ge-detector with random coincidences (\ref{sec:larvetocut}) -- see Table \ref{tab:sources}. Without a source the pulser acceptance is 97.3\%, and the PMT trigger rate ($\sim$5\,kHz) is dominated by dark noise and the decay of $^{39}$Ar (1.4\,kBq).

The LAr veto and PSD cuts are applied to each measurement in the whole energy range. While the achieved suppression factor in the \Roi is the ultimately relevant number, other energy regions of distinct gamma lines illustrate the different and complementary suppression mechanism underlying PSD and LAr veto.

\subsection{Th-228 suppression}

Even after careful material screening and selection, $^{228}$Th and its progenies from the natural decay chain are present at trace levels in the construction materials of \Gerda . Background from sources close-by the germanium originates from detector holders, cables and front-end electronics \cite{Gerda_background}. These components are immersed in the liquid argon and are referred to as `close-by sources'. Conversely, `external sources' are located in the cryostat, its neck and the photomultipliers of the LAr instrumentation itself. The corresponding energy spectra of a close-by and external $^{228}$Th source are shown in Fig. \ref{fig:Th228_full}.

\begin{figure}
	\footnotesize{\input{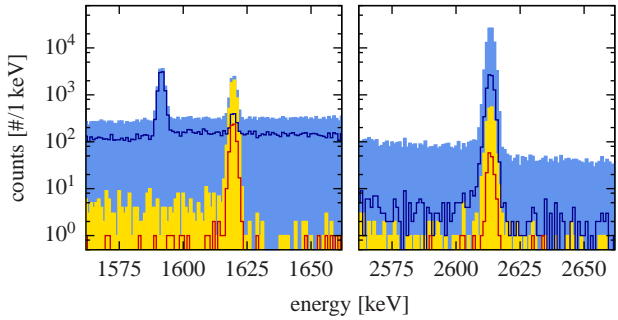}}
	\caption{\textbf{Left:} the double escape peak of $^{208}$Tl at 1593\,keV and the 1621\,keV \emph{single} gamma line of $^{212}$Bi. \textbf{Right:} the 2615\,keV \emph{coincident} gamma line of $^{208}$Tl. Both examples from the close-by source. Same colour code as in Fig. \ref{fig:full}.}
	\label{fig:DEP1621}
\end{figure}

Both spectra are dominated by the 2615\,keV gamma line of $^{208}$Tl and it's single- and double escape peak (SEP at 2104\,keV, DEP at 1593\,keV). Since the double escape peak is dominantely of SSE nature, and as such used to calibrate the PSD acceptance, it remains practically unsuppressed by PSD (Fig. \ref{fig:DEP1621}). On the other hand, the two 511\,keV annihilation gammas trigger the LAr veto reliably such that the peak vanishes. In contrast to the DEP, the neighbouring 1621\,keV full energy peak from the $^{212}$Bi decay is affected nearly opposite by both cuts: all the energy of this line is deposited in the germanium detector, leaving none for the surrounding LAr to create scintillation. Moreover, the transition is not part of a gamma cascade, thus no additional energy deposition in the LAr is occuring. Hence, the LAr veto does not come in. The suppression factor of PSD is about ten for both source positions. The situation is the same for PSD on the 2615\,keV gamma line. However, the LAr veto can also suppress this peak, because the gamma is emitted in a cascade preceded by other gammas (\emph{coincident gammas}), which can themselves trigger the LAr veto. The suppression is much stronger for the close-by source ($SF=47$) than for the external ($SF=1.3$), because in the former case coincident gammas have little chance to escape from the active LAr volume. This instance can be exploited to identify the location of a $^{228}$Th background source via the LAr suppression factors. 

The \Roi of the \znbb-decay is dominated by a flat Compton region of $^{208}$Tl before and after the cuts in both spectra (Fig. \ref{fig:Th228_ROI}). The suppression factors of the LAr veto cleary differ for the close-by ($1180\pm250$) and external ($25\pm1$) sources, while being quasi-independent of the source location for PSD ($2.4\pm0.1$ and $2.8\pm0.1$ respectively -- see Table \ref{tab:SF}). The LAr suppression is strongly enhanced in the Compton region of the 2615\,keV line, as compared to the full energy peak itself: since a fraction of about 2\,MeV is deposited in the Ge-detectors, an excess energy of $\sim$600\,keV is deposited in the liquid argon in its vicinity, providing an additional handle for the LAr veto to act upon. The suppression factors of the combined LAr veto and PSD cuts are $5200\pm1300$ (close-by) and $129\pm15$ (external), thus providing a strong suppression on all $^{228}$Th background sources. Since only 5 counts survive the combined suppression of the close-by source, the analysis window was extended from 70\,keV to 200\,keV width around \Qbb. Still, the number of counts after the cut (=15) dominate the uncertainty of the suppression factor.

\subsection{Ra-226 suppression}

$^{226}$Ra is a long lived progeny of $^{238}$U in the natural decay chain. Similar to $^{228}$Th, it is present at trace levels in all construction materials. $^{226}$Ra decays to $^{222}$Rn, which is a radioactive noble gas that diffuses into the LAr of the \Gerda cryostat and is a potential source of background. Energy spectra were recorded with \LArGe for a close-by and external $^{226}$Ra source, both of which are shown in Fig. \ref{fig:Ra226_full}.

The predominant isotope in this chain that contributes to the region at \Qbb is $^{214}$Bi, with several gamma lines up to 3184\,keV. The suppression of gamma peaks follows the same logic as showcased for $^{228}$Th: \emph{single} lines not beeing affected by the LAr veto (e.g. 1764\,keV, 2204\,keV, 2448\,keV), as opposed to lines emitted as part of a gamma cascade and therefore in \emph{coincidence} (e.g. 609\,keV, 1120\,keV). The LAr suppression factors in the \Roi are $4.6\pm0.2$ (close-by) and $3.2\pm0.2$ (external), and about four for PSD (see Table \ref{tab:SF}, Fig. \ref{fig:Ra226_ROI}). The LAr suppression is much inferior compared to $^{228}$Th for mainly two reasons: (1) all gamma lines with sufficient energy to create Compton events at \Qbb are \emph{single}, hence depriving the LAr veto of a possibility to veto on a \emph{coincident} gamma. The lack of coincident gammas also makes the veto less dependent on the source position. And (2) the gammas have only little energy exceeding \Qbb, thus only little light is created to trigger the veto. Despite the individual suppression of LAr veto and PSD being moderate, their combination again provides a suppression well beyond one order of magnitude.

\subsection{Co-60 suppression}

Cosmogenic $^{60}$Co is formed in Ge-detectors and their copper holders during production above ground. While $^{60}$Co in the detectors can create background directly via their beta decay, $^{60}$Co background from the holders relies on gammas: only two gamma lines with significant branching ratio are emitted after the decay of $^{60}$Co. Since their energies of 1173\,keV and 1332\,keV are below \Qbb, they can create background events only via summation. As summation strongly depends on the solid angle and the angular correlation of the gammas, only $^{60}$Co sources close-by the Ge-detectors are of concern for the \Gerda background. The energy spectrum of close-by $^{60}$Co as measured in \LArGe is shown in Fig. \ref{fig:Co60_full}.

The energy region above the two gamma lines is dominated by the summation spectrum, which expands up to the summation peak at 2505\,keV\footnote{The peak above 2505\,keV is the 2615\,keV line of $^{208}$Tl background from the front-end electronics used with the BEGe detector.}. Similar to $^{214}$Bi, LAr suppression in the \Roi (Fig. \ref{fig:Co60_ROI}) happens mainly via the gamma energy exceeding \Qbb, which in case of $^{60}$Co is almost 500\,keV. This higher energy and multiplicity reduces the chance for it beeing deposited in dead volume rather than active argon, hence leading to a superior suppression factor of $27\pm2$ compared to $^{226}$Ra (Table \ref{tab:SF}). The PSD cut works very efficiently with a suppression of $76\pm9$, as by construction summation events are MSE. Again, the combined cut can reject background by three orders of magnitude. 

\subsection{Conclusion on suppression factors}

\begin{table}
\begin{center}
	\caption{\label{tab:SF} Summary table of suppression factors for different sources in the \Roi around \Qbb at 2039\,keV. The combined suppression (`total') ranges from a few 10$^1$ to several 10$^3$ and gives the order of magnitude by which background sources can be suppressed.}	
  	\begin{tabular}{ccccc}
    \toprule source & position & \multicolumn{3}{c}{suppression factor} \\ & & LAr veto & PSD & total \\
    \midrule $^{60}$Co  				  & int & $27\pm2$   & $76\pm9$  & $3900\pm1300$ \\
    \midrule \multirowbt{2}{*}{$^{226}$Ra}& ext & $3.2\pm0.2$  & $4.4\pm0.4$ & $18\pm3$      \\
    \cmidrule{2-5}                        & int & $4.6\pm0.2$  & $4.1\pm0.2$ & $45\pm5$      \\
    \midrule \multirowbt{2}{*}{$^{228}$Th}& ext & $25\pm1$   & $2.8\pm0.1$ & $129\pm15$    \\
    \cmidrule{2-5} 						  & int & $1180\pm250$ & $2.4\pm0.1$ & $5200\pm1300$ \\
    \bottomrule 
  \end{tabular}
  \vspace{-1.8em}
\end{center}
\end{table}

The suppression factors of the different sources in the \Roi of the \znbb-decay are summarized in Table \ref{tab:SF}. The large variation of the suppression factors is consistent with our understanding of the underlying physics, as described in the previous sections.

The combination of LAr veto and PSD proves to be more powerful than would be expected from independent cuts: the mean average of the combined suppression in the \Roi is enhanced by a factor $1.84\pm0.17$, compared to the product of the individual suppression factors of PSD and LAr veto. This means that event classes leading to rejection by one or the other cut are anti-correlated. For example, a Compton event at \Qbb from close-by $^{208}$Tl leaves the 2.6\,MeV gamma to deposit $\sim$600\,keV outside the Ge-detector, alongside its 583\,keV coincident partner. However, if the event at \Qbb results from the summation of the two gammas (a multisite event likely to be rejected by PSD), a single gamma of higher energy $\sim$1.2\,MeV can leave the detector, and is more likely to escape the active LAr volume than two energetically lower gammas. Hence, this event class is anti-correlated in LAr veto and PSD. An analogue analysis of full energy peaks of the investigated sources returns an average `enhancement' of $1.007\pm0.015$. This is consistent, since no (anti-)correlated event topologies are expected here.

The suppression factors measured here sketch the order of magnitude by which active background rejection in {\Gerda} may be achievable, indicating that the goal for \mbox{Phase II}, namely to reduce the background by one order of magnitude, is in reach. Due to a higher argon purity we anticipate to achieve a better photoelectron yield in {\Gerda}, supporting the suppression of $^{226}$Ra and other background sources that create little light per event at $Q_{\beta\beta}$. Other determining factors for photoelectron yield are the geometry of the active LAr volume, the Ge-detector strings within, and the efficiency of light detectors and wavelength shifters. A full MC campaign using photon tracking has been carried out to study and optimize a veto design for {\Gerda} \mbox{Phase II}. These simulations are in reasonable agreement with the experimental data of \LArGe and will be reported in a publication dedicated to the LAr veto in {\Gerda}.

\section{Background measurements in natural germanium}
\label{sec:background}

\LArGe has been designed to demonstrate the applicability of the LAr veto in an ultra low-background environment. For that purpose a measurement with a semi-coaxial detector with natural isotopic composition (GTF44, see \ref{sec:Germanium detectors}) and improved radiopure front-end electronics was conducted. The LAr veto was operated under the same conditions as previously described, except for an exchange of the LAr by high purity argon 6.0. The veto acceptance of 91\% is determined by the pulser. The detector resolution is 3.5\,keV at 1332\,keV, no PSD was applied.

\begin{figure*}
	\footnotesize{\input{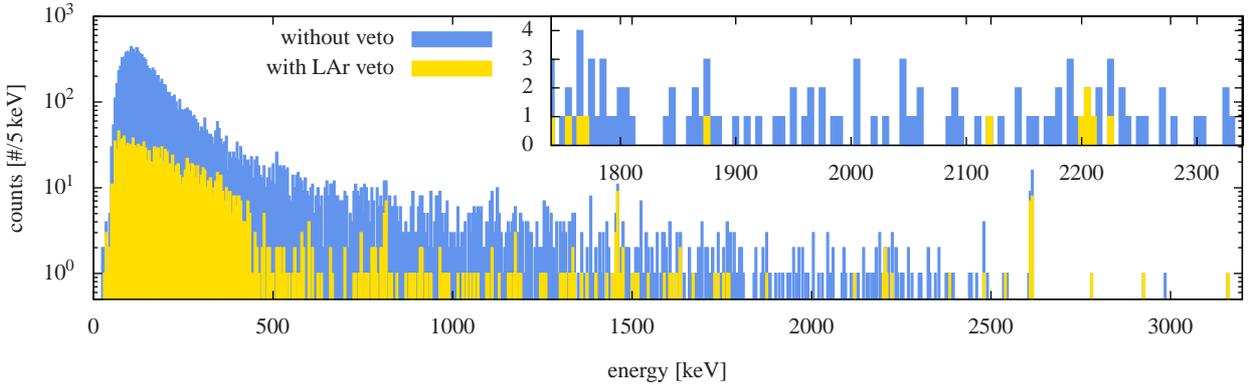}}
	\caption{Background spectrum of GTF44 with an exposure of 116\,kg$\cdot$d (life time 47.05 days). The inlay shows a close-up view of the region of interest around \Qbb at 2039\,keV.}
	\label{fig:bgGTF44}
\end{figure*}

\begin{table}
\begin{center}
	\caption{\label{tab:BG} Summary table of the background index for different choices of the \Roi. All \Roi are centered around \Qbb and exclude the 2204\,keV line of $^{214}$Bi. Confidence intervals are given with 90\%.}
  	\begin{tabular}{ccccc}
    \toprule region & \multicolumn{2}{c}{counts [\#]} & \multicolumn{2}{c}{BI [cts/(keV$\cdot$kg$\cdot$yr)]}
    		\\ of interest & no veto & veto & no veto & veto \\
    \midrule 100\,keV & 14 & 0 & $0.44(12)$ & $<7.2\cdot10^{-2}$ \\
	\midrule 200\,keV & 30 & 1 & $0.47(9)$ & $0.17-6.8\cdot10^{-2}$ \\
	\midrule 300\,keV & 40 & 1 & $0.43(7)$ & $0.12-4.6\cdot10^{-2}$ \\
    \bottomrule 
  \end{tabular}
  \vspace{-1.8em}
\end{center}
\end{table}

\subsection{Background components}

The full energy spectrum with an exposure of 116\,kg$\cdot$d (life time 47.05 days) is shown in Fig. \ref{fig:bgGTF44}. The dominant background source above 1.5\,MeV is the 2615\,keV line of $^{208}$Tl and its Compton continuum. The suppression factor of the full energy peak is $1.52\pm0.62$, which is in agreement with the corresponding value $1.28\pm0.01$ obtained in the external source measurement. Hence, the suppression factor points towards a distant origin of this background, presumably the PMTs. Other prominent background sources are $^{40}$K (1461\,keV) and $^{214}$Bi (1764\,keV and 2204\,keV). Their lines appear only in the vetoed spectrum: while the continuous Compton background is vetoed effectively, single full energy peaks are rejected only by random coincidences. Cosmogenic $^{58}$Co is found at 811\,keV, likely sitting in the activated copper of the detector encapsulation. At low energies the spectrum is dominated by the $^{39}$Ar beta spectrum ($Q_\beta$-value 565\,keV) and accompanying Bremsstrahlung photons.

\subsubsection{\tnbb contribution}

The strong suppression in the vetoed background spectrum makes it possible to observe the \tnbb spectrum even though the detector is made from non-enriched natural germanium. The prediction of the \tnbb spectrum based on \cite{primakoff59,Gerda13_2nbb} for this detector infers that 69 out of 135 observed counts in the continuum from 500\,keV (above dominant bremsstrahlung from $^{39}$Ar) to 2100\,keV (above the \tnbb endpoint) are expected to stem from \tnbb decays.

\subsubsection{$^{42}$K abundance}

A unique background to \Gerda and \LArGe is that of $^{42}$K, identified through its gamma line at 1525\,keV. $^{42}$K ($Q_\beta=3525$\,keV, $T_{1/2}=12.6$\,h) is a $\beta$-emitting progeny of $^{42}$Ar ($Q_\beta=599$\,keV, $T_{1/2}=32.9$\,yr) with traces expected in natural argon. In \LArGe we observe 7 counts in the interval (1523$-$1527)\,keV around the peak, out of which (1.35$\pm$0.27) counts are expected from Compton background. The probability\footnote{The probability is calculated from a gaussian distributed background $g(\lambda|\mu,\sigma)$ and poisson distributed counts $p(n|\lambda)$, using $prob(n$\,$\geq$\,$7)=\sum\nolimits_{n=7}^\infty\int_0^\infty g(\lambda|\mu,\sigma)p(n|\lambda)d\lambda$.} to observe $\geq$7 events from this background is 0.08\%. Neglecting possible inhomogeneities of the $^{42}K$ spatial distribution, the number of counts observed in the 1525\,keV peak corresponds to an abundance $^{42}Ar/^{nat}Ar$ of about \linebreak\mbox{$2\cdot10^{-21}$\,g/g} \cite{Heisel11}. Along with \Gerda \cite{Gerda_background}, the data presented here is the first positive detection of natural $^{42}$Ar. 

\subsection{Background Index}

The inlay of Fig. \ref{fig:bgGTF44} shows the \Roi of the \znbb-decay. In a large energy window of 300\,keV centered around \Qbb only one event survives the LAr veto cut. Depending on the chosen width for that region, the achieved background index after veto is about $10^{-2}$ cts/(keV$\cdot$kg$\cdot$yr) -- see Table \ref{tab:BG}. The lower limits cover a background index of $10^{-2}$ cts/(keV$\cdot$kg$\cdot$yr), which is the design goal of \LArGe and \Gerda \mbox{Phase I}. The 90\% confidence intervals are determined for `the mean of a Poisson variable in the absence of background' using frequentists statistics according to \cite{PDB10}. The ratio of counts before and after the LAr veto yield a background suppression by one order of magnitude or more.

\section{Conclusion}

The \LArGe test facility has demonstrated the great potential of an active liquid argon veto for the suppression of residual background signals which deposit part of their energy in LAr. It is the first time bare Ge-detectors are operated in a low-background environment with 1\,m$^3$ of instrumented LAr. The background suppression efficiency has been studied in combination with pulse shape discrimination (PSD) of a BEGe detector. Suppression factors have been measured for several sources ($^{60}$Co, $^{226}$Ra, $^{228}$Th) representing characteristic background sources to \Gerda in different locations (close-by and external). The strongest suppression factors were obtained for combined LAr veto and PSD for close-by $^{228}$Th ($SF\approx5200$) and $^{60}$Co ($SF\approx3900$). The combined suppression of LAr veto and PSD is mutually enhanced. The particular response of the different suppression methods is a useful tool to understand the location of different backgrounds even in the case of low counting statistics. In a low background measurement without PSD, the LAr veto helped to achieve an excellent background index of \mbox{(0.12$-$4.6)$\cdot 10^{-2}$} \mbox{cts/(keV$\cdot$kg$\cdot$y)} (90\% C.L.). The confidence interval coincides with the background level of \Gerda \mbox{Phase I} <$10^{-2}$ \mbox{cts/(keV$\cdot$kg$\cdot$y)}, despite \LArGe being a much more compact setup. \LArGe has the sensitivity to measure the natural abundance of $^{42}$Ar and the \tnbb-decay in non-enriched germanium. As a consequence of these results, an active liquid argon veto has been developed for \Gerda and will be used in \mbox{Phase II} of the experiment.

\section*{Acknowledgements}

The \LArGe experiment is supported financially by the German Federal Ministry for Education and Research (BMBF), the German Research Foundation (DFG) via the Excellence Cluster Universe and the SFB/TR27, the Russian Foundation for Basic Research (RFBR), and the National Centre for Research and Development (Poland, ERANET-ASPERA/04/11). The institutions acknowledge also internal financial support. The authors thank the staff of Laboratori Nazionali del Gran Sasso for their continuous strong support of the experiment.


\end{document}